# On the multiferroic skyrmion-host GaV$_4$S$_8$

S. Widmann,[a] E. Ruff,[a] A. Günther,[a] H.-A. Krug von Nidda,[a] P. Lunkenheimer,[a] V. Tsurkan,[a,b] S. Bordács,[c] I. Kézsmárki,[a,c,d] and A. Loidl[a*]

[a]Experimental Physics V, Center for Electronic Correlations and Magnetism, University of Augsburg, 86135 Augsburg, Germany
[b]Institute for Applied Physics, Academy of Sciences Moldova, Chisinau MD-2028, Republic of Moldova
[c]MTA-BME Lendület Magneto-optical Spectroscopy Research Group, 1111 Budapest, Hungary
[d]Department of Physics, Budapest University of Technology and Economics, 1111 Budapest, Hungary

*Corresponding author, Email: alois.loidl@physik.uni-augsburg.de

The lacunar spinel GaV$_4$S$_8$ exhibits orbital ordering at 44 K and shows a complex magnetic phase diagram below 12.7 K, which includes ferromagnetic and cycloidal spin order. At low but finite external magnetic fields, Néel-type skyrmions are formed in this material. Skyrmions are whirl-like spin vortices that have received great theoretical interest because of their non-trivial spin topology and that are also considered as basic entities for new data-storage technologies. Interestingly, we found that the orbitally ordered phase shows sizable ferroelectric polarization and that excess spin-driven polarizations appear in all magnetic phases, including the skyrmion-lattice phase. Hence, GaV$_4$S$_8$ shows simultaneous magnetic and polar order and belongs to the class of multiferroics, materials that attracted enormous attention in recent years. Here, we summarize the existing experimental information on the magnetic, electronic, and dielectric properties of GaV$_4$S$_8$. By performing detailed magnetic susceptibility, resistivity, specific heat, and dielectric experiments, we complement the low-temperature phase diagram. Specifically, we show that the low-temperature and low-field ground state of GaV$_4$S$_8$ seems to have a more complex spin configuration than purely collinear ferromagnetic spin order. In addition, at the structural Jahn-Teller transition the magnetic exchange interaction changes from antiferromagnetic to ferromagnetic. We also provide experimental evidence that the vanadium V$_4$ clusters in GaV$_4$S$_8$ can be regarded as molecular units with spin 1/2. However, at high temperatures deviations in the susceptibility show up, indicating that either the magnetic moments of the vanadium atoms fluctuate independently or excited states of the V$_4$ molecule become relevant.

## 1. Introduction

Lacunar spinels are ternary chalcogenides of composition AM$_4$X$_8$ (A = Ga and Ge; M = V, Mo, Nb, and Ta; X = S and Se). They represent an interesting class of transition-metal compounds with molecular clusters as structural building blocks [1,2,3]. Lacunar spinels are derived from the normal spinel structure AM$_2$X$_4$ by removing every second A-site ion. Due to long-range ordered defect formation, the bond lengths are rearranged and, thus, the structure consists of two weakly linked molecular units, cubane (M$_4$X$_4$)$^{n+}$ and tetrahedral (AX$_4$)$^{n-}$ clusters. These compounds are dominated by strong electronic correlations and exhibit exotic ground states, including complex spin order, Mott derived metal-to-insulator transitions, and pressure-induced superconductivity [4,5,6]. During the past decade, $AM_4X_8$ lacunar spinels have attracted much attention due to their interesting and tuneable electronic properties such as d-derived heavy-fermion behavior [7,8], bandwidth-controlled metal-to-insulator transition [9], large negative magnetoresistance [10], a two-dimensional topological insulating state [11], and electric field-induced resistive switching [12,13,14,15]. Very recently, the emergence of orbitally-driven ferroelectricity has been reported for $GeV_4S_8$ [16] and $GaV_4S_8$ [17,18]. THz and high-frequency dielectric spectroscopy on $GaV_4S_8$ show that strong polar fluctuations accompany the JT transition and that indeed orbital order drives the ferroelectric polarization [18]. These compounds are novel magnetoelectric multiferroics, which show spin-driven excess-polarization in the different magnetically ordered states. A further important and interesting aspect of lacunar spinels is the mixed (non-integer) valence shown by most of the metallic M ions. In many cases, mixed-valence compounds tend to undergo charge-order transitions at low temperatures. In lacunar spinels it is believed that the metal ions within the cubane $M_4X_4$ clusters form $M_4$ units with a unique molecular electron density [19]. It seems interesting to check if these electronic molecular structures are stable up to elevated temperatures. There are reports in literature that this may not be the case [20].

The lacunar spinel $GaV_4S_8$ studied in the present work is a magnetic semiconductor with non-centrosymmetric cubic $F\bar{4}3m$ symmetry at room temperature. It consists of a network of weakly coupled $(V_4S_4)^{5+}$ cubane units forming a face-centered cubic (fcc) lattice, separated by $(GaS_4)^{5-}$ tetrahedra. The cubane units comprise vanadium $V_4$ molecular clusters, which can be characterized by a local spin ½ [20]. $GaV_4S_8$ undergoes a cubic to rhombohedral structural phase transition at $T_{JT}$ = 44 K and develops magnetic order below 12.7 K [5,17,21,22]. The low-temperature structure has been described by the space group R3m [5,20]. The structural phase transition has been identified as a Jahn-Teller (JT) transition characterized by an elongation of the $V_4$ tetrahedra of the cubane units along one of the four crystallographic <111> directions [23]. In Ref. 24, utilizing magnetic susceptibility, atomic-force microscopy, and small-angle neutron scattering, a complex magnetic phase diagram of $GaV_4S_8$ was deduced, with a cycloidal (Cyc) phase and a skyrmion-lattice (SkL) state embedded within a ferromagnetic (FM) phase. Skyrmion lattices, periodic arrays of spin vortices, have recently been observed in various magnets with chiral structure [25,26,27,28]. Most interestingly, in $GaV_4S_8$ novel, Néel-type skyrmions have been identified [24]. So far, such type of skyrmion has not been observed in bulk materials, but only in magnetic thin films [29]. By detailed dielectric, pyro-, and magnetocurrent experiments, it has been established that the orbitally ordered phase in $GaV_4S_8$ has sizable ferroelectric polarization [17]. In addition, all magnetic phases show excess polarization, including the SkL phase, which is characterized by magnetic skyrmions dressed with ferroelectric polarization [17]. Hence, the skyrmion host $GaV_4S_8$ is an interesting and specific multiferroic, with orbital-order derived ferroelectricity below the JT transition and a zoo of spin-driven polar phases below the magnetic ordering transition, including skyrmions carrying an electric polarization pattern.

Here we report a detailed study of the magnetic, thermodynamic and electronic properties of $GaV_4S_8$ using magnetic susceptibility, electrical resistivity, and heat-capacity experiments, supplemented by measurements of the dielectric permittivity and ferroelectric polarization. First we discuss indications that the spin S = ½ ground state of the $V_4$ molecular units may not be stable at elevated temperatures. We provide experimental evidence that the electrical resistivity dramatically changes at the JT transition, concomitantly with the sign of the exchange interaction. Concerning the polar properties, in agreement with Ref. [17] we find that dominant ferroelectricity with sizable polarization appears at the JT transition at 44 K, while additional spin-driven ferroelectricity results from FM, as well as from Cyc and skyrmion spin textures. From a careful analysis of the magnetic susceptibility we provide experimental evidence that strictly collinear FM order might not be the low-temperature ground state in $GaV_4S_8$.

## 2. Experimental details

Polycrystalline $GaV_4S_8$ was prepared by solid-state reaction using pure elements of Ga (6N), V (3N), and S (5N). Three subsequent synthesis steps were necessary to obtain complete reaction of the starting materials to form the stoichiometric ternary phase. After each step, phase purity was checked by x-ray powder diffraction. The synthesized polycrystals were used as starting material for single-crystal growth using chemical transport reactions. The growth was performed in closed quartz ampoules at temperatures between 800 and 850 °C utilizing iodine as transport agent. Crushed single crystals were characterized by x-ray diffraction and found to be free of any impurity phases. At room temperature we found the correct $GaMo_4S_8$-type crystal structure with $F\bar{4}3m$ symmetry and a lattice constant $a = 0.966$ nm, in agreement with the findings reported in Ref. 20.

Magnetic measurements were performed with a SQUID magnetometer (Quantum Design MPMS XL) in the temperature range from 1.8 K < T < 400 K and in external magnetic fields up to 5 T. Both, two-point constant-voltage resistivity with a Keithley 6517A electrometer and heat capacity were investigated in a Physical Properties Measurement System (Quantum Design PPMS) for temperatures 1.8 K < T < 300 K. The excitation voltage in the resistivity experiments was 140 mV. In conventional heat-capacity experiments (termed "small-pulse" method in the following), heat pulses leading to a temperature increase of 1-2% of the actual sample temperature were applied for a given time $t_0$. The resulting time-dependent temperature curves were fitted on heating and cooling. In large-pulse experiments, we applied heat pulses leading to a 20% temperature increase for a time $3t_0$. The temperature response of the sample was analyzed by a dual-slope method [30] for heating and cooling, except for the first-order JT transition, where only the heating curve was used. This procedure provides high-resolution data and gives access to details of the heat capacity that are smeared out or even completely missed by the conventional, small-pulse method. To establish a detailed low-temperature phase diagram, the heat capacity was determined as function of temperature and also as function of the external magnetic field.

The dielectric properties of $GaV_4S_8$ were measured at audio and radio frequencies, as well as in the microwave range, at temperatures from 4 K to 300 K. Experiments in the frequency range from 1 Hz to 10 MHz were performed with a frequency-response analyzer (Novocontrol Alpha-Analyzer). The high-frequency experiments (1 MHz - 2.5 GHz) were made with an impedance analyzer (Keysight E4991A) using a coaxial reflectometric technique with the sample capacitor mounted at the end of the coaxial line [31]. To determine the ferroelectric polarization, the pyrocurrent was measured utilizing the Keithley 6517A electrometer as function of temperature between 2 K and 55 K. In addition, we detected the magneto-current at temperatures between 2 K and 14 K and in external fields between 0 and 300 mT. In these experiments, we used platelet-shaped single crystals of 1 mm$^2$ cross section and 0.25 mm thickness, in most cases with the electric field applied along the <111> direction.

## 3. Experimental results and discussion

### 3.1 High-temperature magnetic, electric and thermodynamic properties

Before establishing and discussing the low-temperature ferroelectric and magnetic phase diagram, we will discuss various material properties in detail. Figure 1(a) shows the magnetization M measured on single-crystalline $GaV_4S_8$ in the temperature range between 2 and 70 K. In this experiment, the external magnetic field $\mu_oH = 1$ T was oriented along the crystallographic <111> direction, which corresponds to the magnetic easy axis [24]. Figure 1(b) shows the inverse susceptibility $1/\chi = H/M$ in the full temperature range up to 400 K. Clearly, it does not follow a strict Curie-Weiss (CW) law, but reveals the tendency to saturate towards high temperatures. This behavior could indicate some temperature-independent paramagnetic (PM) contribution, i.e., van-Vleck paramagnetism from excited levels of the molecular $V_4$ clusters. It also could point towards the fact that the excited states of the $V_4$ clusters become thermally occupied resulting in enhanced effective moments at high temperatures. Alternatively, it could even be possible that at high temperatures the $V_4$ units do not longer behave as compact entities with a unique molecular electron density, but decouple into independent local PM moments of single vanadium ions. More detailed investigations to even higher temperatures are necessary to clarify these questions.

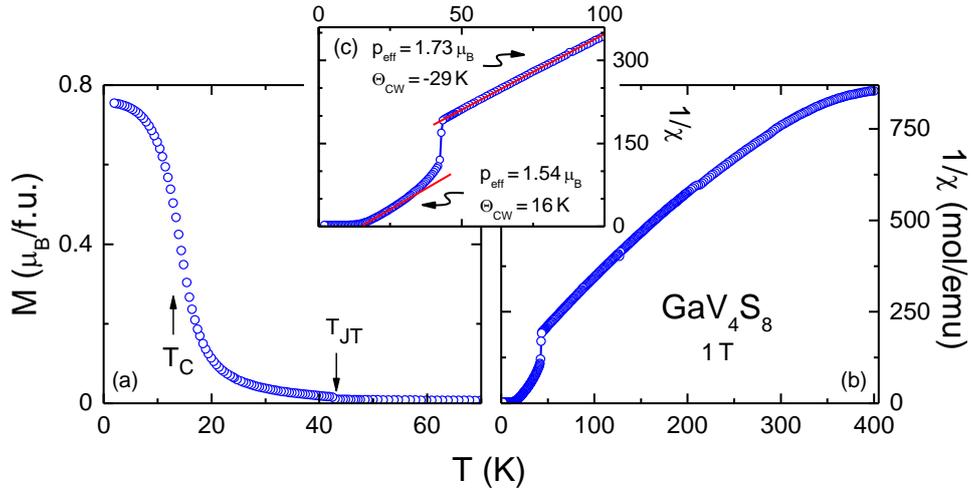

Figure 1. (Color online). (a) Magnetization M from 2 - 70 K measured in an external magnetic field $\mu_oH = 1$ T, indicating the onset of spontaneous FM polarization in $GaV_4S_8$. The structural ($T_{JT} = 44$ K) and magnetic ordering temperatures ($T_C = 12.7$ K) are marked by arrows. (b) Inverse magnetic susceptibility from 2 - 400 K. (c) Enlarged view of inverse susceptibility below 100 K. The CW parameters ($p_{eff} = 1.73$ $\mu_B$, $\theta_{CW} = -29$ K and $p_{eff} = 1.54$ $\mu_B$, $\theta_{CW} = 16$ K) result from CW laws (indicated by lines) in the temperature range from 44 to 100 K and from 18 to 30 K, respectively.

Figure 1(c) demonstrates that the data from 100 K down to the JT transition temperature can well be fitted by a CW law with an effective PM moment $p_{eff} = 1.73$ $\mu_B$ and a negative CW temperature of $\theta = -29$ K. The PM moment is compatible with $S = ½$ of the vanadium molecular entities assuming a g value of 2. The negative CW temperature signals moderate antiferromagnetic exchange in the cubic high-temperature phase above the structural transition. As shown in Figure 1(c), the character of the susceptibility drastically changes below the JT transition at 44 K. A jump-like decrease of the inverse susceptibility, followed by a slightly steeper decrease, indicates FM exchange interactions in the orbitally ordered phase. Despite the fact that, due to the phase transitions, between the two susceptibility anomalies an ideal CW behavior never evolves, a linear approximation between 18 K and 30 K reveals an effective moment of 1.54 $\mu_B$ and a FM CW temperature of 16 K. The former indicates slightly reduced PM moments of the S = ½ $V_4$ molecules, the latter coincides reasonably well with the FM ordering temperature of $GaV_4S_8$. Obviously, at the JT transition, the dominant magnetic exchange turns from antiferromagnetic to FM due to strong local structural distortions. The appearance of spontaneous magnetization below the FM transition temperature $T_c$ = 12.7 K is documented in Figure 1(a). At low temperatures, the ordered moment at 1 T is of the order 0.8 $\mu_B$/f.u., which is close to the value expected for vanadium $V_4$ molecules, carrying in total a spin S = ½. Already at this point, it should be noted that at lower magnetic fields (< 0.2 T) a much more complex magnetic phase diagram evolves, including Cyc and SkL spin order.

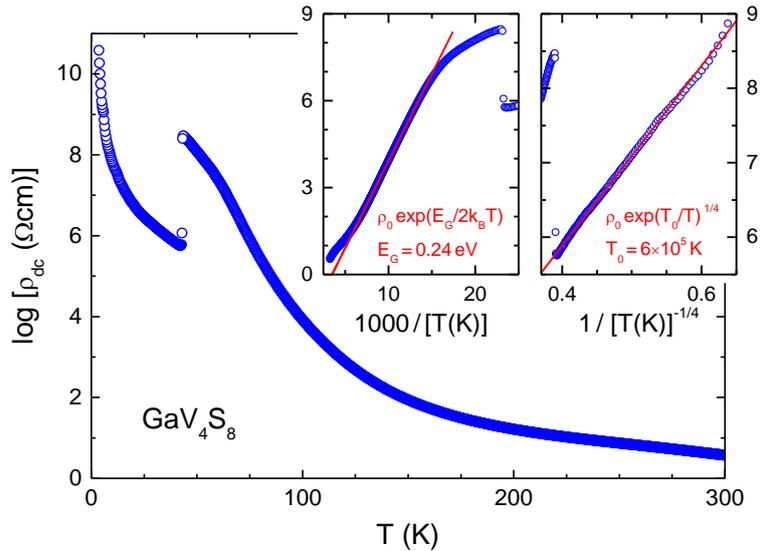

Figure 2. (Color online). Temperature dependence of the electrical resistivity of $GaV_4S_8$ on a semi-logarithmic plot from 2 to 300 K. Left inset: Arrhenius representation of the high-temperature data (T > $T_{JT}$). The solid line corresponds to an Arrhenius fit to determine the relevant electronic band gap energy. Right inset: Log resistivity vs. $T^{-1/4}$ to document VRH conductivity (T < $T_{JT}$). The solid line represents a fit assuming charge transport via a VRH mechanism (see text).

Figure 2 shows the electrical resistivity $\rho_{dc}$ of $GaV_4S_8$ measured in a single-crystalline sample between 2 K and room temperature. For all temperatures, we find semiconducting behavior with an overall increase of the conductivity from 4 $\Omega$cm at room temperature to about $10^{10}$ $\Omega$cm at 4 K. We detected an enormous step-like drop of the resistivity upon cooling the sample through the JT transition by almost three orders of magnitude, which remains the same on heating and cooling, with a small hysteresis of less than 0.5 K. The temperature dependence of the electrical resistivity shows almost no sign of the onset of magnetic order at 12.7 K. To get an estimate of the electronic band gap of this compound, for temperatures above the JT transition, the left inset shows the logarithm of the resistance plotted against the inverse temperature. In this Arrhenius representation, linear behavior can be identified between 70 and 170 K. When assuming a purely intrinsic semiconducting behavior, $\rho_{dc} \sim \exp(E_g/2k_BT)$, the band gap $E_g$ amounts 0.24 eV. Most of the lacunar spinels investigated were characterized as narrow-gap semiconductors with electronic band gaps in the 0.2 - 0.3 eV range [15].

Below the JT transition, we found that purely activated behavior does not seem to be the adequate transport mechanism. Instead, a plot of the logarithm of the resistivity vs. $T^{-1/4}$ provides a better linearization of the experimental results (right inset of Figure 2). This points towards a variable-range hopping (VRH) mechanism, which implies hopping conductivity of localized electronic charge carriers as the dominant charge-transport mechanism [32]. The resistance can be described by $\rho \sim \exp(T_0/T)^{1/4}$ with an exponent parameter $T_0 \approx 6\times10^5$ K (solid line in the right inset of Figure 2). There is a number of reports on the resistivity behavior of polycrystalline $GaV_4S_8$ [20,21,33,34,35]. In polycrystalline samples, the decrease of the resistivity above the JT transition is not as strong as documented in Figure 2 and the step at the JT transition is well below one decade [35]. At room temperature, all these studies find energy barriers of the order of 0.14 eV, which decreases significantly on cooling. Above the JT transition, the conductivity has also been analyzed in terms of VRH [33,34]. As mentioned above, in our single-crystalline sample we find Arrhenius behavior to provide a better description over a fairly broad temperature range, however with significant deviations just above the JT transition. At this point, we would like to recall that lacunar spinels are prime candidates for resistive switching [14,15] and the resistance therefore strongly depends on the voltage used, especially at low temperatures.

Figure 3 shows the temperature dependence of the heat capacity from 2 K up to room temperature. For presentation purposes, we choose a C/T vs log(T) representation. A well-defined and rather narrow lambda-like anomaly appears at the JT transition. It allows locating the structural phase transition with high precision at $T_{JT}$ = 44 K and documents the high quality of the single-crystalline material used in this work. The magnetic phase transition can be identified at 12.7 K. Both anomalies are only weakly shifted on heating and cooling. Based on symmetry considerations, the structural phase transition from the cubic paraelectric to the JT distorted FE state must be a first-order transition. Figure 3 shows results from both small-pulse (conventional) and large-pulse experiments, which should yield identical results. Specifically, both data sets should agree at second-order phase transitions, while strong discrepancies can appear in first order transitions. The results from both experiments coincide in the complete temperature range investigated. Only close to the structural transition, the heat capacity (C/T) measured in the large-pulse modus yields values close to 11 J/mol $K^2$ as compared to 1.5 J/mol $K^2$ in conventional scans. This discrepancy indeed is a direct experimental evidence of the first-order character of the JT transition.

The solid line in Figure 3 represents the phonon contributions to the heat capacity and has been calculated by assuming one Debye term for the three acoustic modes. The contributions of the expected 12 Einstein terms were formally fitted by three modes accounting for the remaining 36 degrees of freedom. A good fit up to room temperature was obtained using a Debye temperature of 204 K and Einstein frequencies corresponding to 240 K, 460 K, and 650 K. To get an estimate of magnetic contributions at the lowest temperatures, in the upper inset of Figure 3 we plot the low-temperature heat capacity as $C/T^{3/2}$ vs $T^{3/2}$, which is a characteristic representation of the heat capacity of isotropic ferromagnets and allows separating magnon and phonon contributions. The linear fit following the Bloch $T^{3/2}$ law is indicated by the solid line, corresponding to $C/T^{3/2} = \gamma + \beta\, T^{3/2}$, where $\gamma$ inversely depends on the spin stiffness and, hence, on the magnetic ordering temperature, while $\beta$ allows calculating the phonon-derived Debye temperature. The best fit yields $\gamma = 36.6$ mJ/mol $K^{5/2}$ and $\beta = 2.98$ mJ/mol $K^4$. Due to the low FM ordering temperature, the magnon contribution is unusually large. The good fit indicates that phonon and magnon excitation alone can describe the low-temperature heat capacity of $GaV_4S_8$ and that an electronic contribution from free charge carriers is not needed. It is also worth mentioning that a pure FM fit works well, even at zero external magnetic field, despite the fact that the ground state just below the ordering temperature is a Cyc phase, but obviously with neighboring spins being close to a collinear configuration. From the prefactor $\beta$ of the phonon term of the heat capacity, a Debye temperature of 204 K can be calculated. This value was used in the calculation of the phonon-derived heat capacity as described above.

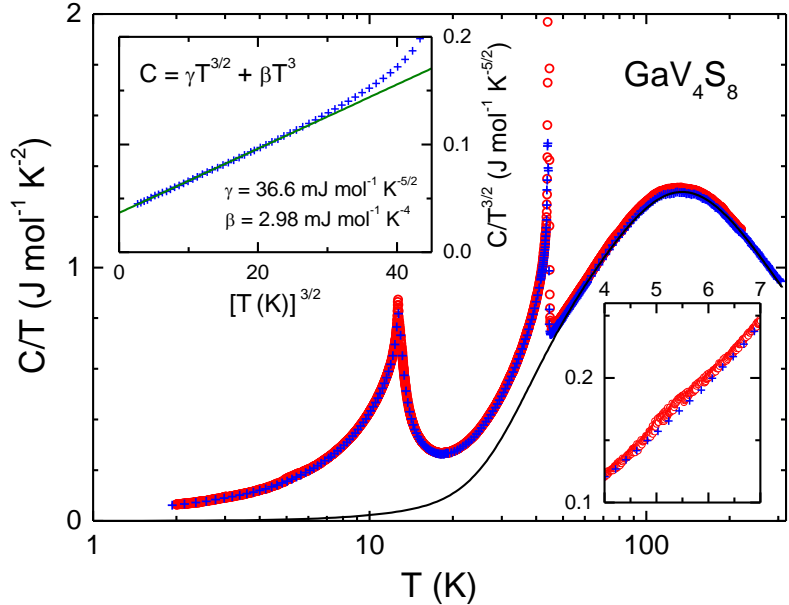

Figure 3. (Color online). Temperature dependence of the heat capacity from 2 K up to room temperature. The representation C/T vs. a logarithmic temperature axis has been chosen for presentation purposes. Blue crosses and open red circles indicate the results of small and large heat-pulse experiments, respectively. Large-pulse experiments yield values of C/T close to 11 J/mol $K^2$ (peak values > 2 J/mol $K^2$ are not shown). The solid line corresponds to a fit approximating the pure phonon contributions (see text). Upper inset: Low-temperature heat capacity plotted as $C/T^{3/2}$ vs $T^{3/2}$ allowing separating magnetic and phononic contributions (see text). Lower inset: Magnified view of the heat capacity around 5 K indicating a small anomaly in the large-pulse experiments. The anomaly is time dependent pointing towards a metastable low-temperature phase.

By utilizing the large-pulse technique, we were able to identify a small anomaly in the heat capacity close to 5 K. This is shown in the lower inset of Figure 3 on an enlarged scale. This anomaly seems to be time dependent, depends on heating or cooling rates, and therefore obviously does not indicate a transition between stable thermodynamic phases. As will be outlined later, it probably corresponds to a transition from the Cyc state to a spin state with short-range order only. This state, however, should be very close to a collinear FM spin configuration.

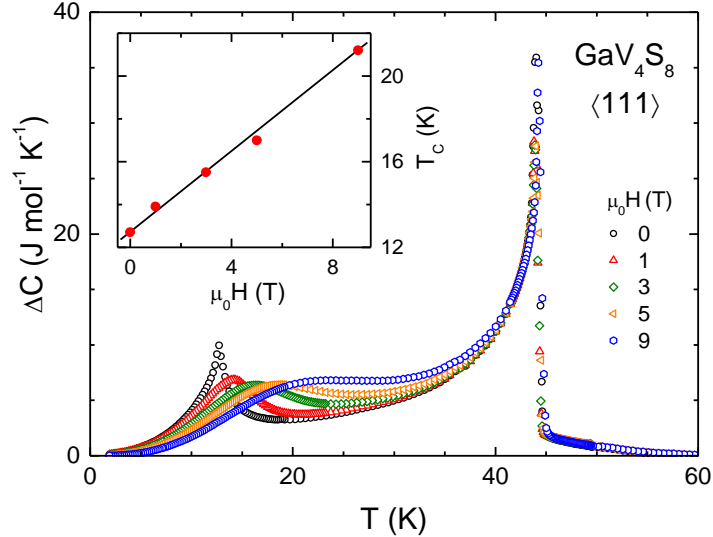

Figure 4. (Color online). Magnetic-field dependence of the heat capacity after subtracting the phonon contributions, which are indicated by the solid line in Figure 3. The excess heat-capacity ΔC at magnetic fields between 0 and 9 T is plotted vs. temperature. The inset shows the dependence of the peak maximum of the magnetic anomaly vs. the external magnetic field. The solid line is drawn to guide the eye.

We also investigated the heat capacity of $GaV_4S_8$ at external magnetic fields up to 9 T. Figure 4 presents the excess heat-capacity after subtracting the phonon derived contributions as indicated by the solid line in Figure 3. The structural anomaly at the JT transition is almost field independent. In contrast, the FM transition reveals a strong shift to higher temperatures and is considerably smeared out as usually observed for field-induced transitions. The field dependence of the maxima of the observed anomalies is plotted in the inset of Figure 4. As a collinear spin state is favored by an external magnetic field, at 9 T the FM transition is shifted to about 21 K, almost enhanced by 100% compared to the zero-field result.

To characterize the polar properties of $GaV_4S_8$, Figure 5 shows the temperature dependence of the real part of the dielectric permittivity ε' and the real part of the complex ac conductivity σ' for frequencies between 1 MHz and 2.5 GHz as function of temperature between 10 K and 70 K. One should note that σ'(T) also reflects the behavior of the imaginary part of the permittivity ε''(T), due to the relation $\sigma' = \omega\varepsilon''\varepsilon_0$ (ω: angular frequency, $\varepsilon_0$: permittivity of vacuum). For an analysis of the dielectric properties, we focus on high frequencies (> 1 MHz) only. As discussed in detail in the Supplemental Material of Ref. [17], the low-frequency dielectric response is dramatically influenced by extrinsic Maxwell-Wagner like contributions, which arise from contact and surface effects and yield ε' values of order $10^4$ at room temperature [36,37]. In the temperature dependence of the dielectric constant and of the conductivity significant anomalies are observed at the JT transition. As revealed by Figure 5(a), at $T_{JT}$ the real part of the dielectric permittivity at 1 MHz exhibits a peak-like anomaly with a continuous increase and decrease under cooling above and below the transition, respectively, typical for a ferroelectric transition. This dielectric anomaly is strongly frequency dependent and becomes increasingly suppressed at higher frequencies, resembling the typical behavior of order-disorder ferroelectrics [38]. However, one should also have in mind that the transition is strongly of first order, and the ferroelectricity certainly has to be characterized as improper, namely driven by the onset of orbital order.

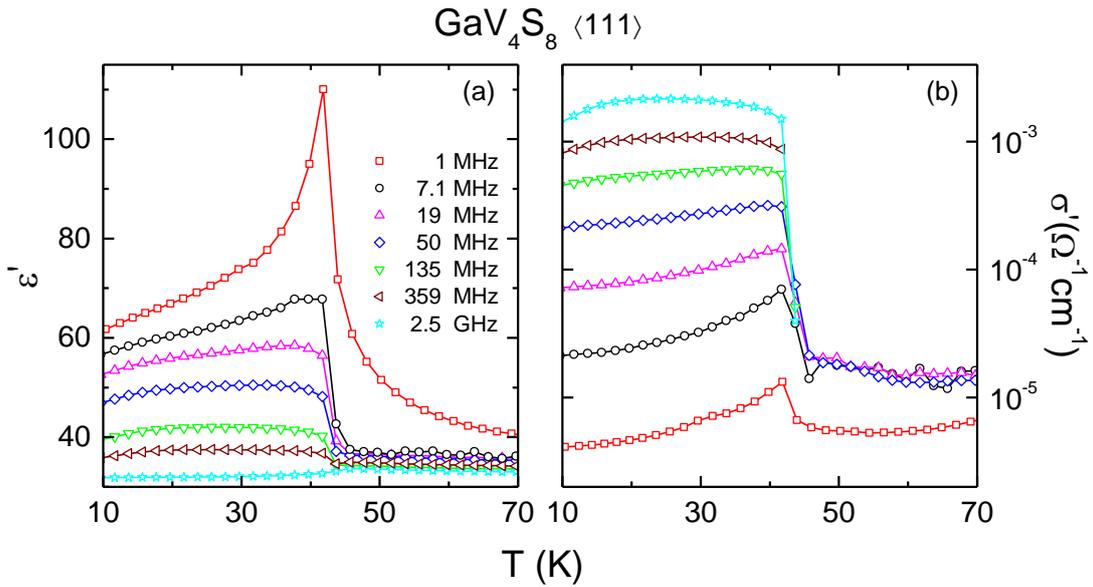

Figure 5. (Color online). Temperature dependence of the dielectric properties of $GaV_4S_8$ with the electric field parallel to the crystallographic <111> direction. (a) Real part of the dielectric constant between 10 and 70 K for frequencies between 1 MHz and 2.5 GHz. (b) Real part of the conductivity in a semilogarithmic plot in the same parameter range of temperatures and frequencies. In both frames the lines are drawn to guide the eye.

The real part of the conductivity (Figure 5(b)) shows a step-like increase at the JT transition. This step again is strongly frequency dependent and amounts almost two orders of magnitude at MHz frequencies. It has been shown that the unusual temperature and frequency dependence of the complex dielectric constant as documented in Figure 5 results from a dramatic change of the relaxation dynamics at the first-order JT transition [18], where the mean dipolar relaxation time slows down by orders of magnitude. Above the JT transition the characteristic frequencies of the relaxation dynamics of coupled orbital and polar degrees of freedom are in the THz regime and only little dispersion effects are visible at audio and radio frequencies. In contrast, in the orbitally ordered state the relaxations are in the low-frequency regime, as is documented in Figure 5 and discussed in detail in Ref. [18].

The well-pronounced anomalies of the dielectric permittivity at the structural transition contrast with the almost invisible effects at the FM transition. The onset of magnetic order is hardly visible in the temperature dependence of $\varepsilon'$, only revealing a small hump close to 13 K, and does not show up in the temperature dependence of the conductivity. This correlates nicely with the fact that in the dc resistivity (Figure 2) the onset of magnetic order can hardly be identified, too.

Finally, Figure 6 shows the electric polarization as determined via pyrocurrent measurements with the electric field parallel to the crystallographic <111> direction. The ferroelectric polarization abruptly appears in a narrow temperature range just below the structural transition. The ferroelectric order parameter (i.e., the polarization) rises in a temperature range as narrow as the width of the transition determined by heat capacity measurements (Figure 3). Hence this transition has to be characterized as a structural first-order transition where concomitantly orbital order and ferroelectric polarization are established. The polarization saturates already close to 40 K, at values above 0.5 µC/cm$^2$. This is significantly larger than the polarization in spin-driven multiferroics [39], only by a factor of 50 lower than in canonical perovskite ferroelectrics, and close to the polarization found in the related lacunar spinel GeV$_4$S$_8$ [16].

Orbital order and ferroelectricity are usually fully decoupled phenomena. Only very recently JT distortions have been identified as a novel source of ferroelectricity [40,41] or even of multiferroicity [42]. Obviously, lacunar spinels belong to this rare class of systems showing orbital-order driven ferroelectricity. Based on a general group-theory analysis, it was concluded that the non-centrosymmetric nature of the high-temperature phase of GaV$_4$S$_8$ is the origin of the ferroelectric JT transition [41].

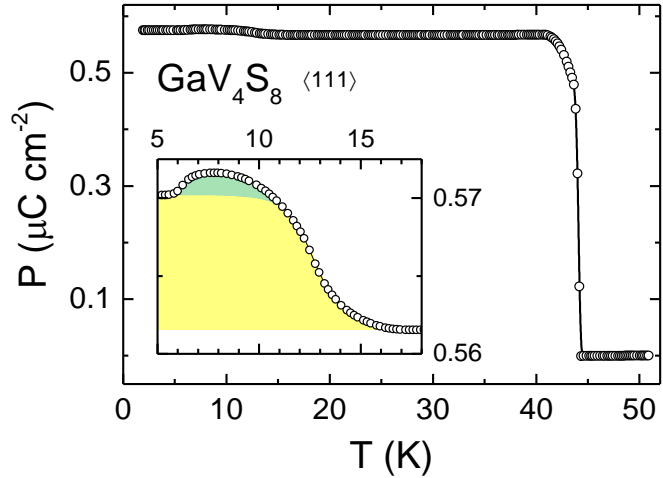

Figure 6. (Color online). Temperature dependence of the polarization in GaV$_4$S$_8$ along the crystallographic <111> direction. The inset provides an enlarged view of the polarization at low temperatures around the magnetic phase transitions. The polarization due to almost collinear spin order is indicated by the yellow area. The spin-driven excess polarization of the Cyc phase, which exists in a narrow temperature range between 6 K and T$_C$ is given by the green area.

A closer inspection of Figure 6 reveals an additional small step in the temperature dependence of the polarization at the onset of magnetic order. The inset of Figure 6 provides an enlarged view and reveals an increase of the polarization when entering the Cyc phase at 12.7 K and a small subsequent decrease close to 6 K, when finally collinear FM spin order is established (however, as shown below, short-range Cyc or SkL order may still play a role). This documents that in GaV$_4$S$_8$, in addition to orbital-order induced ferroelectric polarization, spin-driven polarization exists in both magnetic phases. It should be noted that this spin-derived polarization is a sub-percent effect compared to the polarization due to orbital order. This will be discussed in the following paragraph in more detail.

### 3.2 Low-temperature magnetic and polar phase diagram

To construct a low-temperature magnetic phase diagram, we have performed systematic magnetization measurements between 2 K and 13 K, which for the relevant regions of the phase diagram were taken in steps of 0.25 K. Typical scans with the external magnetic field along the crystallographic <111> direction are documented in Figure 7(a). Similar experiments have been performed with the magnetic field along the main crystallographic directions <110> and <100> as well. At almost all temperatures investigated, the magnetization shows field-induced upturns, indicative for metamagnetic phase transitions.

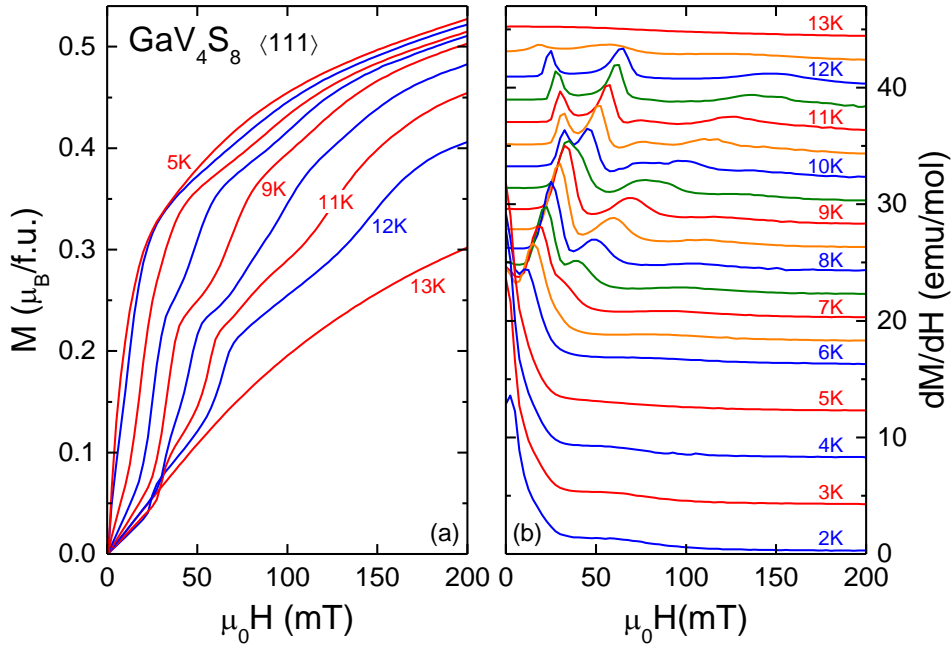

Figure 7. (Color online). (a) Low-temperature magnetization of $GaV_4S_8$ as measured in external magnetic fields along the <111> direction and between 0 and 200 mT for a series of temperatures between 5 and 13 K. (b) Derivative of the magnetization for external magnetic fields between 0 and 200 mT and temperatures between 2 and 13 K. For clarity, the absolute values of the derivatives have been shifted along the vertical axis.

To construct a detailed phase diagram, Figure 7(b) shows the derivatives of some representative M(H) scans. In the following, we use the peak maxima of the derivatives to determine characteristic phase-transition temperatures. At low temperatures (< 6 K), the main structure in the derivatives of the magnetization is a peak close to zero external field resulting from domain and demagnetization effects. An additional weak shoulder, probably indicating minor spin reorientations, is observed close to 60 mT at T < 6 K. It successively shifts to higher fields for higher temperatures. At 6 K, in addition a strong low-field peak evolves at finite fields, indicating the transition from the Cyc phase into the collinear FM state [24]. Above approximately 10 K, this peak splits into a well-pronounced two-peak structure signifying the opening of a phase pocket, which is characteristic for the SkL phase. In addition, at 7 K a shoulder in dM/dH(H) arises (at about 35 mT), developing into a broad peak that shifts to higher fields with increasing temperature and also splits into two smeared-out peaks above about 9 K. Thus, for example at 10 K, in addition to the mentioned dominating peak pair enclosing the SkL phase, there are three additional broad peaks (at about 75, 100, and 136 mT) signaling a complex phase diagram. This can be partly explained by the strong easy-axis anisotropy active in $GaV_4S_8$, where along <111> only one domain out of four is aligned parallel to the external field. In contrast, the rhombohedral easy axes in the other three domains span an angle of 71° with respect to the external field. Hence, in these domains the spin structures survive up to higher fields. The spin structures in these domains, which are misaligned with respect to the external magnetic field, are denoted by stars (Cyc*, SkL*) in the following. It has been impressively documented before that all the phase boundaries due to the different domains in $GaV_4S_8$ can be mapped onto one unique phase diagram after the magnetic field axis is scaled by the direction cosine between the easy axis and the external field [24]. The phase boundaries between the Cyc* and SkL* phases, as well as between SkL* and FM obviously are not characterized by strong anomalies in the field-dependent magnetization but only show up as relatively weak and broad peaks in the derivatives dM/dH (Figure 7(b)).

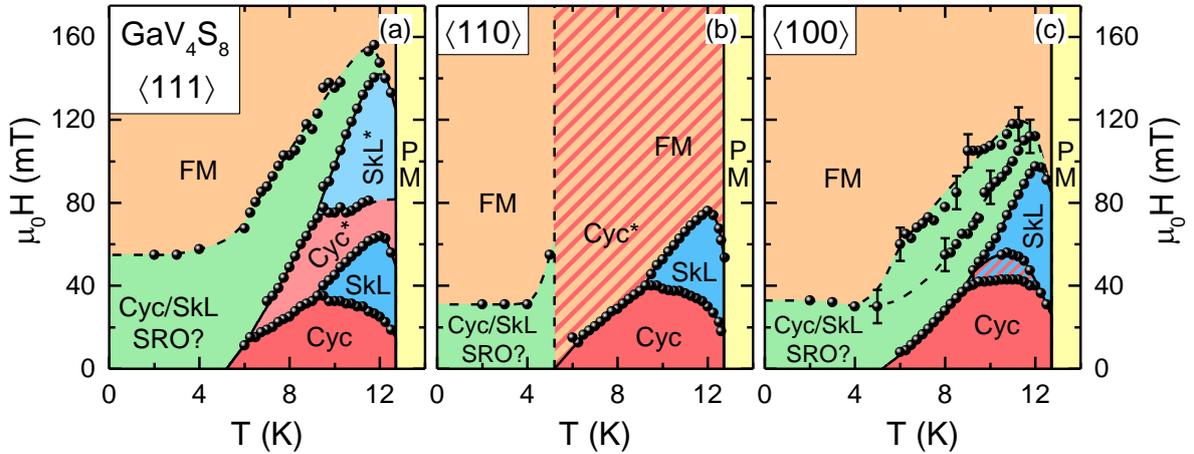

Figure 8. (Color online). Low-temperature magnetic phase diagram of $GaV_4S_8$ along the three principle crystallographic directions. The Cyc, Cyc*, SkL, and SkL* phases are indicated (see text). The phases shown in green probably indicate SRO regimes of either Cyc or skyrmion phases. Along <110>, two domains are perpendicular to the external field and the Cyc* phase survives in these domains up to the spin-flip transition (dashed area). All these domains are embedded between the PM and the field-induced FM order. For temperatures T > 12.7 K and at low fields, PM behavior has been identified.

Based on these careful magnetization measurements, we reconstructed the low-temperature magnetic phase diagram of $GaV_4S_8$ for the three principal directions of a cubic crystal. The results are shown in Figure 8 and resemble the main results as published in Ref.

[24]: Cyc and SkL phases are embedded between the PM and the field-induced FM phases. They represent the response of domains which are aligned parallel or have small angles with respect to the external field while the Cyc* and SkL* phases result from domains spanning significant angles with the field. In these domains, the Cyc and SkL spin patterns survive up to significantly higher fields. Despite the fact that the overall phase diagram is similar to that published by Kézsmárki et al. [24], differences become apparent. Towards low temperatures or high fields, the Cyc and SkL phases are followed by smeared-out transitions (Figure 7(b)) pointing to additional phases, which are indicated by green areas in Figure 8. They are not simple collinear FM phases, but most likely are characterized by short-range Cyc or SkL order. Figures 8(a) and (c) document for the <111> and <100> directions that neither Cyc nor SkL phases have direct phase boundaries with the FM collinear phases, but are separated by these complex short-range ordered (SRO) or phase-separated regions. Detailed neutron-scattering experiments are necessary to elucidate these phase regimes. The situation is quite different for the external field along the <110> direction (Figure 8(b)). Here two domains are perpendicular to the external field. These two domains probably survive with Cyc spin order up to fields of 0.4 T [43], which is indicated by the hatched area in Figure 8(b). In this area no further phase boundaries were detected. Along <100>, a small and narrow phase pocket between the Cyc and SkL phase and a second phase boundary within the SRO regime appear, whose nature at present cannot be explained and certainly deserves further studies. It cannot be excluded that these findings result from that the fact that not all domains exactly span an angle of 55° with the external field.

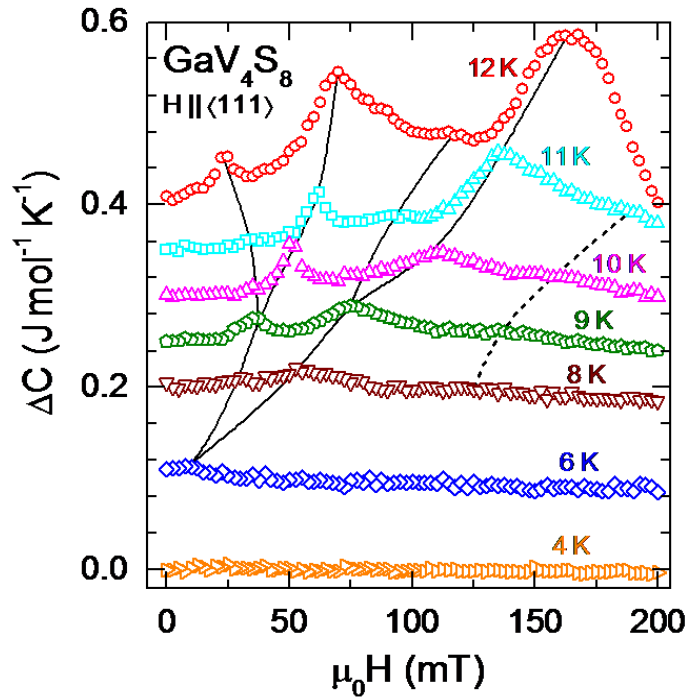

Figure 9. (Color online). Results of magnetocaloric experiments on $GaV_4S_8$ with the external magnetic field applied along the <111> direction: Heat capacity as function of the external magnetic field between 0 and 200 mT for a series of temperatures between 4 and 12 K. Tentative phase boundaries are indicated by solid and dashed lines.

The phase diagrams obtained from magnetization measurements, as documented in Figure 8, can nicely be mirrored by heat-capacity experiments, which have been conducted as function of the external magnetic field for temperatures between 4 and 12 K. The results for the crystallographic <111> direction are shown in Figure 9. Well-defined peaks appear at the phase boundaries between the SkL and Cyc* and between the SkL* and FM (or SRO) phase boundaries. These peaks signal significant entropy changes on passing the phase boundaries. The transitions between Cyc to SkL phase reveal minor anomalies only, while the boundaries between Cyc* and SkL* phases are identified by weak shoulders only. The phase boundary between the SRO phases and the induced FM phase are visible as broad smeared-out anomalies only between 8 and 11 K. We indicate the development of the transition temperatures in Figure 9 by solid or dashed lines. They are roughly compatible with the phases as indicated in Figure 8(a). The transition between the SRO state and the FM collinear phase below 6 K does not appear in the magnetocaloric experiments. Obviously, it is characterized by minimal entropy changes only. We would like to recall that the transition between the Cyc and the FM phase at zero temperature appeared as a weak time-dependent anomaly only (see lower inset of Figure 3).

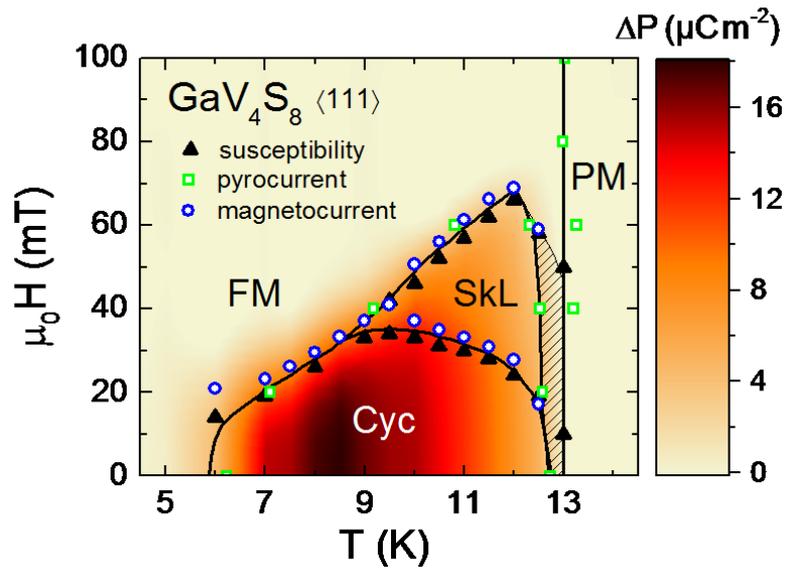

Figure 10. (Color online). Color-coded intensity contour plot of the excess polarization in $GaV_4S_8$ as determined by pyrocurrent and magnetocurrent experiments along the crystallographic <111> direction. The polarization in excess to that of the FM or PM background is shown. Please note that this polarization background is temperature dependent and slightly increases from 13 to 5 K (cf. inset of Figure 6). Phase boundaries between the Cyc, SkL, FM, and PM phases are shown by lines. Characteristic temperatures as determined by magnetic susceptibility, pyrocurrent, and magnetocurrent experiments are indicated separately (see legend). The hatched area indicates a possible precursor state to the SkL phase.

Finally, the low-temperature phase diagram has been mapped by a series of pyrocurrent and magnetocurrent experiments. The resulting excess polarization in the Cyc and the SkL phase is shown in Figure 10. Please note that the polarization of the simple FM collinear phase, which is indicated as light-yellow background, represents the zero value in this figure. Its excess polarization amounts about 85 μC/m$^2$ [17], which is almost by a factor of 100 lower than the ferroelectric polarization induced by orbital order (cf. Figure 6). The excess polarizations of the Cyc and SkL phases are nicely documented and different in the different magnetic phases: Figure 10 provides experimental evidence that the spin-driven polarization in GaV$_4$S$_8$ is strongest in the Cyc state, slightly less in the SkL phase, and again slightly smaller in the FM phase. The excess polarization in the Cyc phase, compared to the polarization of the FM phase with strictly collinear spin alignment, amounts 16 μC/m$^2$ and that in the SkL phase approximately 8 μC/m$^2$. The distinct polarization values evidence that the SkL phase carries a finite ferroelectric polarization, implying whirl-like spin textures with polar moments. In Ref. [17] it has been shown that all spin-driven excess polarization can be explained in terms of exchange striction and that the ferroelectric polarization of the skyrmions is donut shaped with a ring of higher polarization around the skyrmion core.

It also should be noted that in the data shown in Figure 10 we observe no excess polarization in the Cyc* or SkL* phases, although the experiments were performed along the crystallographic <111> direction. We speculate that under the given experimental conditions with small and thin platelets of the samples contacted with silver paint, finally single domain samples were formed with the bulk of the magnetic domains aligned along <111> and hence along the field direction.

## 4. Summary and conclusions

In this study we have presented a detailed characterization of the Néel-type skyrmion host GaV$_4$S$_8$. We reported the temperature dependence of the magnetic susceptibility and of the electrical resistivity, provided a detailed analysis of the heat capacity in external magnetic fields up to 9 T, and studied the temperature dependence of the dielectric constant, of the ac conductivity, as well as of the electrical polarization. Based on these systematic studies, we revisited the low-temperature multiferroic phase diagram. The main results of this work can be summarized as follows: i) Towards high temperatures the magnetic susceptibility shows saturation effects, which could indicate that above room temperature the V$_4$ clusters do not have a unique molecular electronic density. ii) Analyzing the inverse susceptibility, we find clear evidence for antiferromagnetic exchange above the JT transition, which changes into FM exchange below. Obviously, the local distortions of the lacunar spinel structure are responsible for this considerable impact on the magnetic exchange. Above (T < 100 K) and below the JT transition the PM moment resembles that of a spin-½ system with a g-value close to or slightly below two. iii) The electrical resistivity reveals semiconducting behavior, characterizing GaV$_4$S$_8$ as narrow-gap semiconductor. For T > T$_{JT}$, the electrical resistivity follows thermally activated behavior with an energy barrier of 0.24 eV. At the JT transition, the resistivity is reduced by three orders of magnitude under cooling. Below the JT transition we found evidence for hopping conduction of localized charge carriers, which can be parameterized in terms of a VRH mechanism. iv) The structural as well as the magnetic transitions are characterized by well-defined anomalies in the temperature dependence of the specific heat. We parametrized the phonon contributions by one Debye and three Einstein modes. The low-temperature specific heat is dominated by magnon contributions and can be well described by the Bloch T$^{3/2}$ law. The unusually large magnon contribution can be explained by the low magnetic ordering temperature. v) Under external magnetic fields, the JT transition remains unchanged while the FM transition is shifted to 21 K in 9 T. vi) The temperature dependence of the complex dielectric permittivity shows clear anomalies at the structural phase transition, which are attributed to the onset of improper ferroelectricity driven by the onset of orbital order. The polarization indicates the first-order character of the transition and reaches sizable values of the ferroelectric polarization already below 40 K. vii) Finally, the low-temperature multiferroic phases were characterized. In addition to the phases established in Refs [17,24], we found indications of short-range magnetic order at low temperatures and low fields, indicating that the ground state of $GaV_4S_8$ probably is more complex and not simply collinear ferromagnetic. We also have characterized the magnetic phases by detailed pyrocurrent and magnetocurrent experiments and derived a color-coded three-dimensional phase diagram documenting the excess polarization which appears in the Cyc and in the SkL phase. The experiments reveal that $GaV_4S_8$ is a genuine multiferroic with spin-driven excess polarization in all magnetic phases and, in addition, with spin vortices that carry ferroelectric polarization. As shown in Ref. [17], all spin-driven polarization stems from an exchange-striction mechanism, which explains the magnitude of the excess polarization of the FM, Cyc, and skyrmion phases.

**Acknowledgements**

This research was supported by the DFG via the Transregional Research Collaboration TRR 80: From Electronic Correlations to Functionality (Augsburg/Munich/Stuttgart). S. B. and I. K. were supported by the Hungarian Research Funds OTKA K 108918, PD 111756 and Bolyai 00565/14/11.